\documentclass[submission,copyright,creativecommons]{eptcs}
\providecommand{\event}{ACL2 Workshop 2023} % Name of the event you are submitting to

\usepackage{iftex,hyperref}
\usepackage[cmex10]{amsmath}
\usepackage{array}
\usepackage{url}
\usepackage{times}
\usepackage{cleveref}
\usepackage{lstautogobble}
\usepackage[dvipsnames,table]{xcolor}
\usepackage{dsfont}
\usepackage[T1]{fontenc}
\usepackage[scaled]{beramono}

\newcommand{\ie}{\emph{i.e.}}
\newcommand{\eg}{\emph{e.g.}}

\newcommand{\R}[0]{\ensuremath{\mathcal{R}}}

\usepackage{makecell}

\newcommand{\gs}{GossipSub}

\newcommand{\fc}{FileCoin}
\newcommand{\Eth}{Eth2.0}

\newcommand{\score}{score function}

\newcommand{\acls}{ACL2s}
\newcommand{\golang}{Golang}

\newcommand{\fin}{\textbf{F}}
\newcommand{\glb}{\textbf{G}}

\newtheorem{property}{Property}
\newcommand{\ag}[1]{$\mathrm{AG}_{#1}$}

\makeatletter
\let\sv@thm\@thm
\def\@thm{\let\indent\relax\sv@thm}
\makeatother

\usepackage{tikz}
\usetikzlibrary{positioning, shapes.geometric, backgrounds}
\usepackage{tikzscale}

\usepackage{mathtools}

\usepackage{enumerate}
\usepackage[shortlabels]{enumitem}

\usepackage{physics}
\usepackage{tikz}
\usepackage{mathdots}
\usepackage{yhmath}
\usepackage{cancel}
\usepackage{color}
\usepackage{siunitx}
\usepackage{array}
\usepackage{multirow}
\usepackage{amssymb}
\usepackage{gensymb}
\usepackage{tabularx}
\usepackage{extarrows}
\usepackage{booktabs}
\usetikzlibrary{fadings}
\usetikzlibrary{patterns}
\usetikzlibrary{shadows.blur}
\usetikzlibrary{shapes}

\usepackage{adjustbox}
\usepackage{float}
\usepackage{pgfplots}
\pgfplotsset{compat=1.8}

\usepackage{derivative}

\newcommand*{\funcfont}{\fontfamily{lmss}\selectfont}
\newcommand*{\codefont}{\ttfamily\small}

\DeclareTextFontCommand{\funcfontify}{\funcfont}
\DeclareTextFontCommand{\codefontify}{\codefont}
\newcommand{\codify}[1]{\ensuremath{\mbox{\codefontify{#1}}}}

\lstdefinelanguage{acl2s}{
  keywords={defdata,definecd,definec,defun,defund,lambda,property,record,list,cons,match,alistof,enum,listof,alistof,range,sig,mget,mset,map,map*,reduce,reduce*},
  alsoletter={-},
  otherkeywords={:hyps,:name,:fixed-vars,defdata-alias,defdata-subtype,create-map*,create-reduce*,check=},
  comment=[l]{;;}
}

\ifpdf
  \usepackage{underscore}         % Only needed if you use pdflatex.
  \usepackage[T1]{fontenc}        % Recommended with pdflatex
\else
  \usepackage{breakurl}           % Not needed if you use pdflatex only.
\fi

\title{Verification of GossipSub in ACL2s}
\author{
Ankit Kumar \qquad Max von Hippel \qquad Panagiotis Manolios \qquad Cristina Nita-Rotaru
\institute{Northeastern University\\
Boston, USA}
\email{ \{kumar.anki, vonhippel.m, p.manolios, c.nitarotaru\}@northeastern.edu}
}

\newcommand{\titlerunning}{Verification of GossipSub in ACL2s}
\newcommand{\authorrunning}{A. Kumar, M. von Hippel, P. Manolios \&
  C. Nita-Rotaru}

\hypersetup{
  bookmarksnumbered,
  pdftitle    = {\titlerunning},
  pdfauthor   = {\authorrunning},
  pdfsubject  = {EPTCS},               % Consider adding a more appropriate subject or description
}
\begin{document}
\maketitle

\begin{abstract}
  \gs\ is a popular new peer-to-peer network protocol designed to
  disseminate messages quickly and efficiently by allowing peers to
  forward the full content of messages only to a dynamically selected
  subset of their neighboring peers (mesh neighbors) while gossiping
  about messages they have seen with the rest. Peers decide which of
  their neighbors to graft or prune from their mesh locally and
  periodically using a score for each neighbor. Scores are calculated
  using a \score\ that depends on mesh-specific parameters, weights
  and counters relating to a peer's performance in the network. Since
  a GossipSub network's performance ultimately depends on the
  performance of its peers, an important question arises: Is the score
  calculation mechanism effective in weeding out non-performing or
  even intentionally misbehaving peers from meshes?  We answered this
  question in the negative in our companion
  paper~\cite{kumar2022formal} by reasoning about \gs\ using our
  formal, official and executable ACL2s model. Based on our findings,
  we synthesized and simulated attacks against \gs\ which were
  confirmed by the developers of \gs, \fc, and \Eth, and publicly
  disclosed in MITRE CVE-2022-47547. In this paper, we present a
  detailed description of our model. We discuss design decisions,
  security properties of \gs, reasoning about the security properties
  in context of our model, attack generation and lessons we learnt
  when writing it.
\end{abstract}

\section{Introduction}
\gs\ is a new peer-to-peer network protocol used by popular
applications like \Eth~\cite{EthWhitepaper} and
FileCoin~\cite{FCwhitepaper}. Messages transmitted in a \gs\ network
are typically categorized into \emph{topics}, which peers of the
network can subscribe to or unsubscribe from. A peer can be part of several
meshes corresponding to different topics. In contrast to \emph{flood
  publishing} where a peer forwards every message it receives to all
of its neighboring peers subscribed to the corresponding topic, \gs\
uses \emph{lazy pull}, wherein a peer forwards full messages only to
its \emph{mesh neighbors} in the relevant topic. A peer can graft or
prune a mesh neighbor based on various heuristic security mechanisms
that ultimately rely on a locally computed score.  The score is
calculated periodically by each peer for each of its neighbors and is
never shared. The \score, which is used to calculate a neighboring
peer's score, depends on application-specific parameters and weights,
and takes into account the performance of the neighbor both generally
and on a given topic. Ideally, the score of misbehaving peers (\eg,
peers that drop messages or forward invalid ones) is penalized, which
matters because negatively scored mesh neighbors get pruned.

The \gs\ developers specified their protocol in English
prose~\cite{gsreadme,gs1.0,gs1.1} and implemented it in
GoLang~\cite{golanggs}. They relied on unit-tests and network
emulation~\cite{vyzovitis2020gossipsub,leastAuth20} of pre-programmed
scenarios for testing to show that misbehaving peers in a \gs\ network
are eventually pruned. However, simple testing is not enough. Dijkstra
famously quoted: ``Program testing can be a very effective way to show
the presence of bugs, but it is hopelessly inadequate for showing
their absence.''

In our companion paper~\cite{kumar2022formal}, we formalized the \gs\
specification in the ACL2s (the ACL2
Sedan)~\cite{DillingerMVM07,acl2s11} theorem prover. ACL2s extends
ACL2~\cite{acl2,acl2-web} with an advanced data definition framework
(\emph{Defdata})~\cite{defdata}, the
\emph{cgen}~\cite{chamarthi-integrating-testing,cgen,harsh-fmcad,harsh-dissertation}
framework for automatic counterexample generation, a powerful
termination analysis based on calling context graphs~\cite{ccg} and
ordinals~\cite{ManoliosVroon03, ManoliosVroon04, MV05}, and a
property-based modelling/analysis framework, each of which helped
immensely in our formalization and verification effort. Our publicly
available model~\cite{gsmodel} is designed to be modular and pluggable
\ie, it allows us to reason about parts of the model in isolation as
well as about applications running on top of the network. Officially,
\gs\ does not come with any properties. We formalized and attempted to
prove security properties which (1) we thought should be reasonable
for a score-based protocol like \gs\ to satisfy and which (2) the \gs\
developers agree with. One such property, which states that
continuously misbehaving peers are eventually pruned, turned out to be
invalid in the case of \Eth. We leveraged the cgen facility in ACL2s
to automatically discover vulnerable network states that invalidate
our property. We built attack gadgets using sequences of events which
can take a network from a reasonable starting state to one of the
discovered vulnerable states. We synthesized attacks which were
confirmed by the \gs, \fc, and \Eth\ developers and publicly disclosed
in MITRE CVE-2022-47547. 
At the time of writing, the \gs\ developers are actively working on a fix.
These results are explained in our companion
paper~\cite{kumar2022formal}. In this work, we focus on our
formalization of the \gs\ ACL2s model and its use for reasoning,
simulation and attack synthesis.

\noindent
\textbf{Paper Outline.} Section~\ref{sec:model} describes the \gs\
protocol and our ACL2s model
simultaneously. Section~\ref{sec:reasoning} describes properties we
used to reason about \gs. Based on the insights gleaned from
proving/disproving these properties, Section~\ref{sec:attack}
describes how we synthesized attacks that can disrupt communication in
an \Eth\ \gs\ network. Section~\ref{sec:limitations} describes
some limitations of our model. Section~\ref{sec:related} presents related work
on mechanized theorem proving efforts in the field of distributed
systems. Section~\ref{sec:conclusion} concludes.

\section{GossipSub Model Description}
\label{sec:model}
\begin{figure}[hbt]
  \centering
  \includegraphics[width=0.8\linewidth]{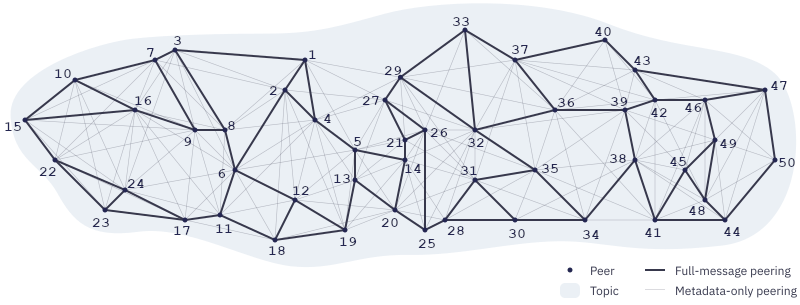}
  \caption{A mesh of peers subscribing to the same topic.  This Figure was taken from~\cite{libp2p-overview}.}
  \label{fig:mesh}
\end{figure}

\begin{figure}
  \centering
  \includegraphics[width=0.5\linewidth]{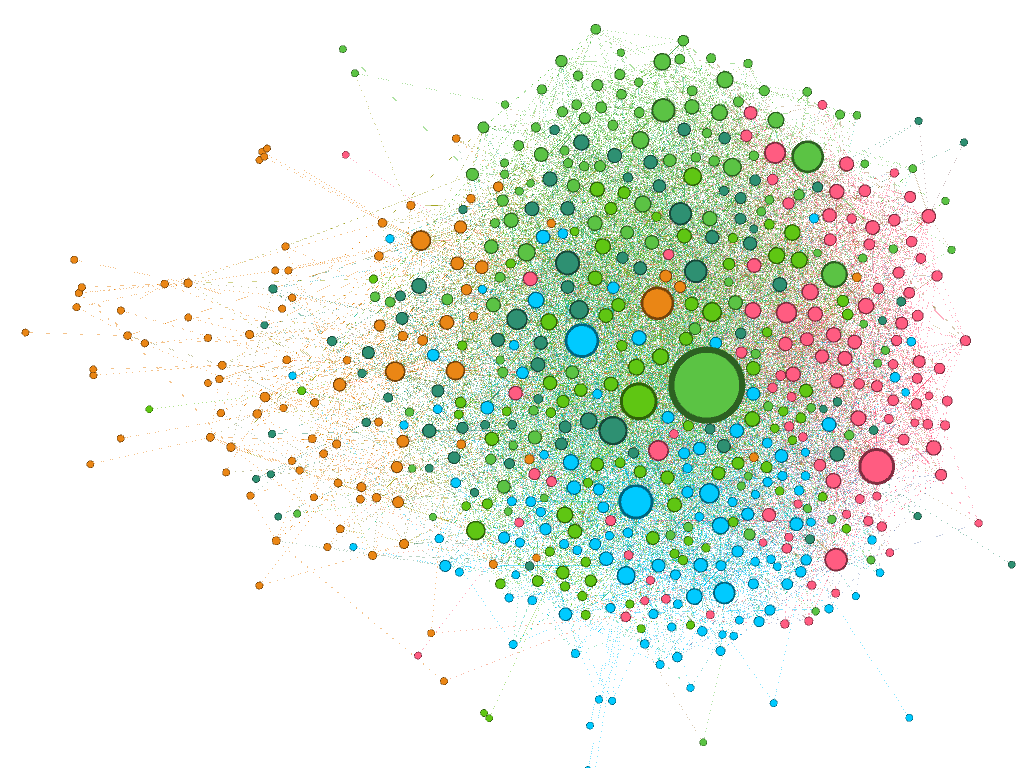}
  \caption{A community graph of \Eth\ nodes where a bigger node
    implies greater degree. Similarly colored nodes have more edges
    among them than to the rest of the graph.}
  \label{fig:network}
\end{figure}

In this section, we describe our ACL2s model, while also giving an
overview of how the \gs\ protocol works. Interested readers are
encouraged to walk through our ACL2s formalization whose presentation
mirrors this section~\cite{demo}. Consider the mesh shown in
Figure~\ref{fig:mesh}.
Full-message payloads are forwarded on the full-message peering edges,
which consume more bandwidth, while the metadata-only peering edges
are used only to advertise and request full-messages using
corresponding ``\codify{IHAVE}'' and ``\codify{IWANT}'' Remote
Procedure Calls (RPCs). These RPCs carry the metadata of the
full-message being advertised or requested, which is considerably
smaller than the message itself. In this way, network congestion is
kept in check, and full-messages are supposed to be eventually
disseminated to all peers who subscribe to the given topic.  As an
example, peers numbered \codify{1} and \codify{2} can send
full-messages to each other, since they are mesh neighbors on the
illustrated topic and share a full-message peering edge. However, peer
\codify{29} can receive a full-message from \codify{1} only if it
requests one, by sending an \codify{IWANT} message in response to a
corresponding \codify{IHAVE} message received from peer \codify{1}.
Note, Figure~\ref{fig:mesh} only shows a single \gs\ mesh. However,
real world applications can have an arbitrary number of topics and
corresponding meshes, \eg, \Eth\ supports up to 71
topics. Figure~\ref{fig:network} illustrates a community graph of
an actual \Eth\ network from Li et. al.~\cite{toposhot}.

In order to model and analyze \gs\ in ACL2s, we rely heavily on
Defdata to easily model network components, and on cgen for property-based testing. 
Distributed systems are often described in terms
of \emph{state machines}, namely automata that encode the discrete
behaviors of a peer in the system~\cite{Lam78, Sch90}.  In our model,
we describe the state-space of a \gs\ peer with a \codify{peer-state}
type, and then implement the \gs\ state machine using a state
transition function. We use a Defdata map from peers to their
corresponding states to capture the state of an entire network.
\begin{lstlisting}
(defdata group (map peer peer-state))
\end{lstlisting}
We use records whenever we need to store state because of the
convenience of using named fields to access the internals of the
state. While proving theorems referring to states, we noticed that we
routinely had to prove helper lemmas about the types of the contents
of those states. This motivated us to rewrite records in the ACL2s
books to enable such helper lemmas automatically, leading
to cleaner code.

The local state of a peer is modeled as a record using the following definition:
\begin{lstlisting}
(defdata peer-state
  (record (nts        . nbr-topic-state)
          (mst        . msgs-state)
          (nbr-tctrs  . pt-tctrs-map)
          (nbr-gctrs  . p-gctrs-map)
          (nbr-scores . peer-rational-map)))
\end{lstlisting}
where (1) \codify{nts} is a record that stores information about the
peer's neighbors' subscriptions, the peer's mesh neighbors, and the
peer's fanout (which we describe later); (2) \codify{mst} is a record
that stores the peer's messages and related state; (3)
\codify{nbr-tctrs} and \codify{nbr-gctrs} are total maps that store
counters used for computing a neighbor's topic-specific and general
scores (respectively); and finally (4) \codify{nbr-scores} is a total
map from peers to their scores. Note that \codify{peer} and
\codify{topic} are ACL2 symbols, while \codify{tctrs} is a record that
keeps track of a peer's behaviors in each topic:

\begin{lstlisting}
(defdata tctrs
  (record (invalidMessageDeliveries . non-neg-rational)
          (meshMessageDeliveries    . non-neg-rational)
          (meshTime                 . non-neg-rational)
          (firstMessageDeliveries   . non-neg-rational)
          (meshFailurePenalty       . non-neg-rational)))
\end{lstlisting}
        
Similarly, \codify{gctrs} keeps track of a peer's general behaviors, not
pertaining to any single topic:

\begin{lstlisting}
(defdata gctrs
  (record (apco . rational)  ;; application provided score
          (ipco . non-neg-rational) ;; ip-colocation factor
          (bhvo . non-neg-rational))) ;; track misbehavior
\end{lstlisting}

Notice that each of the counters is a rational, not a natural. This is
because counters are supposed to fractionally decay at regular
intervals. The rate of decay is dependent on the application running
on top of \gs.  We define maps for each of these counters:

\begin{lstlisting}
(defdata pt (cons peer topic))
(defdata pt-tctrs-map (map pt tctrs))
(defdata p-gctrs-map (map peer gctrs))
(defdata peer-rational-map (map peer rational))
\end{lstlisting}

For each map, we define a lookup function with default values.  
\acls\ automatically proves termination and input-output contracts,
which suffices to show that the maps are total.
As an example, the following is the lookup function for scores:
\begin{lstlisting}
(definecd lookup-score (p :peer prmap :peer-rational-map) :rational
  (let ((x (mget p prmap)))
    (match x
      (nil 0)  ;; default value
      (& x))))
\end{lstlisting}

We now explore each component of the \codify{peer-state}.\\

\noindent
\textbf{Neighbor topics state.} A key objective of the \gs\
protocol is to reduce network congestion, while forwarding data as
quickly as possible. To achieve this, a peer forwards full messages
only to a subset of its neighboring peers. A \gs\ peer keeps
track of the topics its neighbors subscribe to, using the
\codify{nbr-topicsubs} field in \codify{nbr-topic-state}. Using this
information, it is able to build up a picture of the topics around it
and which peers are subscribed to each topic. The peers to which it
forwards full messages in topics it does not itself subscribe to constitute
a list called a \emph{fanout}, and is stored in the \codify{topic-fanout} field,
which is a map from topics to lists of peers (\codify{lop}). The peer
forwards full messages to other peers with whom it shares mesh membership. 
Mesh memberships are stored in the \codify{topic-lop-map}
field \codify{topic-mesh}. Finally, the peer stores the last time it
published a message to its fanout. This is used to expire a peer's
fanout, if it has been too long since the peer last published.

\begin{lstlisting}
(defdata peer symbol)
(defdata lop (listof peer))
(defdata topic-lop-map (map topic lop))
(defdata topic-nnr-map (map topic non-neg-rational))

(defdata nbr-topic-state
  (record (nbr-topicsubs . topic-lop-map)
          (topic-fanout  . topic-lop-map)
          (last-pub      . topic-nnr-map) 
          (topic-mesh    . topic-lop-map)))
\end{lstlisting}

\noindent
\textbf{Messages state.} \codify{msgs-state} is a record type used to
store information about messages received, requested, seen, or
forwarded by a peer. First we define the types \codify{pid-type},
which is just an alias for the type \codify{symbol}, and the record
type \codify{payload-type}.

\begin{lstlisting}
(defdata-alias pid-type symbol)
(defdata payload-type (record (content . symbol)
                              (pid     . pid-type)
                              (top     . topic)
                              (origin  . peer))) 
\end{lstlisting}

\codify{payload-type} is a record used to represent full messages, carrying
the message content, payload id, the topic of this message and the
peer who originated it. pid-type represents payload ids,
 a hash of a message payload which can identify the full
message content. The msgs-state is defined as follows:

\begin{lstlisting}
(defdata msg-peer (v (cons payload-type peer)
                     (cons pid-type peer)))
(defdata msgpeer-rat (map msg-peer rational))
(defdata msgs-waiting-for (map pid-type peer))
(defdata mcache (alistof payload-type peer))

(defdata msgs-state
  (record (recently-seen   . msgpeer-rat)
          (pld-cache       . mcache)
          (hwindows        . lon) 
          (waitingfor      . msgs-waiting-for) 
          (served          . msgpeer-rat) 
          (ihaves-received . nat)
          (ihaves-sent     . nat)))
\end{lstlisting}

\codify{msgs-state} stores (1) \codify{recently-seen}, a map from
either a full-message or a message hash to the time since
receipt; (2) \codify{pld-cache}, an association
list of full messages and their senders; (3) \codify{hwindows}, or
history windows, a list of naturals where each is the
number of messages received in an interval; (4) \codify{waitingfor}, a
map from message ids of messages that haven't been received yet, to
peers one has sent corresponding \codify{IWANT} requests to; and
(5) \codify{served}, a map from
either a full-message or a message hash to a rational, denoting the
count of the number of times this message was served. The \codify{served} map
helps to
detect peers sending too many \codify{IWANT} messages.
Finally, \codify{msgs-state} contains (6)
\codify{ihaves-received} and \codify{ihaves-sent}, which store the number of
\codify{IHAVE} messages received and sent, respectively.\\

\noindent
\textbf{Events.} Our model includes events that can occur in a
network. Events include peers sending or receiving (1) control
messages like \codify{GRAFT} or \codify{PRUNE} for mesh control,
\codify{SUB}, \codify{UNSUB}, \codify{JOIN} and \codify{LEAVE} for
topics, or \codify{CONNECT1} or \codify{CONNECT2} messages to edit
neighbors; (2) \codify{PAYLOAD} for carrying message payload, or
\codify{IHAVE} for advertising or \codify{IWANT} for requesting
message payloads; and (3) \codify{HBM} for heart-beat events occurring
at each peer at regular intervals. A list of events will have type
\codify{loev}. Events are defined as follows:
\begin{lstlisting}
(defdata verb (enum '(SND RCV)))
(defdata rpc (v (list 'CONNECT1 lot)
                (list 'CONNECT2 lot)
                (list 'PRUNE topic)
                (list 'GRAFT topic)
                (list 'SUB topic)
                (list 'UNSUB topic)))
(defdata data (v (list 'IHAVE lopid)
                 (list 'IWANT lopid)
                 (list 'PAYLOAD payload-type)))
(defdata mssg (v rpc data))
(defdata evnt (v (cons peer (cons verb (cons peer mssg)))
                 (list peer 'JOIN topic)
                 (list peer 'LEAVE topic)
                 (list peer verb peer 'CONNECT1 lot)
                 (list peer verb peer 'CONNECT2 lot)
                 (list peer 'HBM pos-rational)
                 (list peer 'APP payload-type)))
(defdata hbm-evnt (list peer 'HBM pos-rational))
(defdata-subtype hbm-evnt evnt)

(defdata loev (listof evnt))
\end{lstlisting}

Heart-beat events occur at regular intervals at each peer in a \gs\
network. Several maintenance activities are performed during each such
event. Scores are updated for neighboring peers, which are then used
to update mesh memberships. Counters are multiplied by some
application-specific decay factors.  Neighboring peers that are not
part of any mesh are sent \codify{GRAFT} messages at regular
intervals, provided that their addition could improve the average
score of peers in the corresponding mesh. Several of these actions
depend on application-specific weights (used for calculating scores)
and parameters.\\

\noindent
\textbf{Parameters and Scoring.} We store the weights and parameters
relevant for scoring in a record called a \codify{twp}. 
This record totally captures the application-specific configuration
of a \gs\ instance, \eg, we can uniquely specify the configuration used by \Eth\
in a \codify{twp}.  Thus, in order to simulate an application running on top of
\gs\ with our model, all we need to know is the application-specific \codify{twp}.
This is what we mean when we say our model is ``pluggable''.

\begin{lstlisting}
(defdata weights
  (record (w1  . non-neg-rational)
          (w2  . non-neg-rational)
          (w3  . non-pos-rational)
          (w3b . non-pos-rational)
          (w4  . neg-rational)
          (w5  . non-neg-rational)
          (w6  . neg-rational)
          (w7  . neg-rational)))

(defdata params
  (record (activationWindow               . nat)
          (meshTimeQuantum                . pos)
          (p2cap                          . nat)
          (timeQuantaInMeshCap            . nat)
          (meshMessageDeliveriesCap       . pos-rational)
          (meshMessageDeliveriesThreshold . pos-rational)
          (topiccap                       . rational)
          (grayListThreshold              . rational)
          (d                              . nat)
          (dlow                           . nat)
          (dhigh                          . nat)
          (dlazy                          . nat)
          (hbmInterval                    . pos-rational)
          (fanoutTTL                      . pos-rational)
          (mcacheLen                      . pos)
          (mcacheGsp                      . non-neg-rational)
          (seenTTL                        . non-neg-rational)
          (opportunisticGraftThreshold    . non-neg-rational)
          (topicWeight                    . non-neg-rational)
          (meshMessageDeliveriesDecay     . frac)
          (firstMessageDeliveriesDecay    . frac)
          (behaviourPenaltyDecay          . frac)
          (meshFailurePenaltyDecay        . frac)
          (invalidMessageDeliveriesDecay  . frac)
          (decayToZero                    . frac)
          (decayInterval                  . pos-rational)))

(defdata wp (cons weights params))
(defdata twp (map topic wp))
\end{lstlisting}

Note that \gs\ required weights \codify{w3} and \codify{w3b} to be
negative. However, the use of Defdata allowed us to automatically find
that \fc\ used an invalid \codify{twp} because it set \codify{w3} and
\codify{w3b} to zero. We discussed this with the \gs\ developers who
then agreed to allow zero values.  Given \codify{T}, a set of topics
which the neighbor subscribes to; \codify{tctrs}, the neighbor's topic
specific counters; \codify{gctrs}, the neighbor's global counters; and
a \codify{twp} containing entries for each topic our neighbor
subscribes to, the \score\ calculates a neighbor's score as shown
below.

\[
\begin{aligned}
 \mathit{score}(q)
 & = \min\big(\text{TC},\sum_{\tau \in T} \text{tw}^\tau \times
    \sum_{i \in \{1,2,3,3b,4\}} w_i^\tau P_i^\tau \big) +
    w_5 P_5 +
    w_6 P_6 +
    w_7 P_7
\end{aligned}
\]
\emph{where}
\begin{align*}
  \codify{(weights . params)} &= \codify{(mget $\tau$ twp)}\\
  w_i^\tau &= \codify{(weights-w\emph{i} weights)}\\
  P_1^\tau &= \codify{(calcP1}\codify{(tctrs-meshTime tctrs) (params-meshTimeQuantum params)}\\
     &\quad\ \codify{(params-timeQuantaInMeshCap params))} \\
  P_2^\tau &= \codify{(calcP2}\codify{(tctrs-firstMessageDeliveries tctrs) (params-p2cap params))}\\
  P_3^\tau &= \codify{(calcP3}\codify{(tctrs-meshTime tctrs) (params-activationWindow params)} \\
                     &\quad\ \codify{(tctrs-meshMessageDeliveries tctrs)}\\
                      &\quad\ \codify{(params-meshMessageDeliveriesCap params)}\\
                     &\quad\ \codify{(params-meshMessageDeliveriesThreshold params))}\\
  P_{3b}^\tau &= \codify{(calcP3b}\codify{(tctrs-meshTime tctrs) (params-activationWindow params)}\\
                              &\quad\ \codify{(tctrs-meshFailurePenalty tctrs)}\\
                              &\quad\ \codify{(tctrs-meshMessageDeliveries tctrs)}\\
                      &\quad\ \codify{(params-meshMessageDeliveriesCap
                        params)}\\
  &\quad\ \codify{(params-meshMessageDeliveriesThreshold params))}\\
  P_4^\tau &= \codify{(calcP4}\codify{(tctrs-invalidMessageDeliveries
         tctrs))} \\
  P_5 &= \codify{(gctrs-apco gctrs)} \\
  P_6 &= \codify{(gctrs-ipco gctrs)} \\
  P_7 &= \codify{(calcP7 (gctrs-bhvo gctrs))}\\
  \text{tw}^\tau &= \codify{(params-topicweight params)}\\
  \text{TC} &= \codify{(params-topiccap (cddar twp))}
\end{align*}
TC does not depend on any topic, but since it is stored in a
\codify{twp} which is indexed by \codify{topic}, its value is
replicated in each of the corresponding \codify{params}. So, it is
fine to extract its value from the first entry of a \codify{twp}. Note
that for score calculations, we require a non-empty \codify{twp}.

Each of the \codify{calcPi} functions where
$\codify{i} \in \{\codify{1,2,3,3b,4,7}\}$ is used to calculate
contributions to the score by one or more of counter values from
\codify{tctrs}. \codify{calcP1} calculates the contribution to a neighbor's
score based on the time spent in common meshes. \codify{calcP2} awards score
for being one of the first few to forward a message. \codify{calcP3}
calculates penalties due to the mesh message transmission rate being
below a given threshold of
\codify{(params-meshMessageDeliveriesThreshold params)}. Whenever a
peer is pruned, its corresponding \codify{tctrs} counter
\codify{meshFailurePenalty} is augmented by the mesh message
transmission rate deficit. This counter is not cleared even after the
peer has been pruned. \codify{calcP3b} scores mesh message delivery failures
based on the value of this counter. Hence, $P^\tau_{3b}$ is a ``sticky''
value which is supposed to discourage a peer that was pruned because
of under-delivery from quickly getting re-grafted in a mesh. \codify{calcP4}
calculates the penalty on score due to sending invalid
messages. \codify{calcP7} calculates penalties due to several kinds of
misbehaviors described by the \gs\ specification. An example of such
misbehaviors includes spamming with too many \codify{IHAVE} messages
which are either bogus and/or not following up to the corresponding
\codify{IWANT} requests.

\noindent
\textbf{The Transition Function.} We define a transition function
\codify{run-network}, which, given an initial \codify{Group} state and
a list of \codify{evnt}, produces a trace of type \codify{egl}, which
is an alist of \codify{evnt} and \codify{group}. Hence, after running
a simulation, we have access to the state of the \codify{Group} after
each \codify{evnt} was processed. \codify{run-network} depends on the
transition function for the \codify{peer-state} (\codify{transition}),
which depends on the transition functions for \codify{nbr-topic-state}
(\codify{update-nbr-topic-state}) and for \codify{msgs-state}\\
(\codify{update-msgs-state}). For brevity, we mention only the
signatures of each of these functions below. Notice that each of these
signatures represents neatly and concisely the types of the formal
arguments and the function return type, which is very useful in a
large code base.

\begin{lstlisting}
(defdata egl (alistof evnt group)) ;; simulation trace

(definecd run-network (gr :group evnts :loev i :nat r :twp s :nat) :egl
...)

(defdata peer-state-ret
  (record (pst . peer-state)
          (evs . loev)))

(definecd transition
  (self :peer pstate :peer-state evnt :evnt r :twp s :nat) :peer-state-ret
...)

(defdata msgs-state-ret
  (record (mst . msgs-state)
          (evs . loev)
          (tcm . pt-tctrs-map)
          (gcm . p-gctrs-map)))

(definecd update-msgs-state (mst :msgs-state evnt :evnt pcm :pt-tctrs-map
                             gcm :p-gctrs-map r :twp) :msgs-state-ret
...)

(defdata nbr-topic-state-ret
  (record (nts . nbr-topic-state)
          (evs . loev)
          (tcm . pt-tctrs-map)
          (gcm . p-gctrs-map)
          (sc  . peer-rational-map)))

(definecd update-nbr-topic-state (nts :nbr-topic-state
                                  nbr-scores :peer-rational-map
                                  tcm :pt-tctrs-map gcm :p-gctrs-map
                                  evnt :evnt r :twp s :nat) :nbr-topic-state-ret
...)
\end{lstlisting}

A \gs\ peer can select a random subset of its fanout and promote them
as mesh members. It can also selects a random subset of its neighbors
to advertise with ``IHAVE'' messages. Such non-determinism is handled
by sending a random seed \codify{s} as a formal parameter to
\codify{run-network}, which is then propagated to the other transition
functions it depends on. Observe that a single event like full message
forwarding can trigger several more forwards, causing a cascade of
events. Such events are represented by the \codify{evs} field in the
return types of \codify{update-msgs-state} and
\codify{update-nbr-topic-state}. In order to limit the total number of
events processed, we send a natural number \codify{i} as a formal
parameter to the \codify{run-network} function.

When proving contract theorems for the transition functions, we needed
to prove the types of terms returned by utility functions, like
\codify{shuffle}, or ACL2 functions like
\codify{set-difference-equal}. For this, we used polymorphism and
automated type-based reasoning provided by Defdata, as shown below:
 
\begin{lstlisting}
(sig set-difference-equal ((listof :a) (listof :a)) => (listof :a))
(sig shuffle ((listof :a) nat) => (listof :a))
\end{lstlisting}

We made heavy use of higher-order macros written by
Manolios~\cite{acl2sources} to improve the readability of our
code. For example, \codify{create-map*} is a list functor. Given an
admitted function name or a lambda expression \codify{f} of type
\codify{a $\rightarrow$ b}, \codify{create-map*} defines a function
\codify{map*-*f} of type \codify{(listof a) $\rightarrow$ (listof
  b)}. In the following code snippet, we show theorems proving that it
obeys the functor laws, and give an example of its usage. \codify{map*} is a
syntactic sugar that maps function \codify{f} onto a list without referring to the
generated function name \codify{map*-*f}.
\begin{lstlisting}
(definec id (x :all) :all
  x)
;; Proof that create -map* is a list functor
;; 1) functor mapping preserves the identity function  
(property functor-id (xs :tl)
  (== (map* id xs) ;; id is an identity function for lists
      (id xs)))
              
;; 2) functor mapping preserves function composition. f and g are declared
;; using defstub. gof is defined as a composition of g and f
(property functor-comp (xs :tl)
  (== (map* gof xs)
      (map* g (map* f xs))))              

;; function to create a list of SND GRAFT events from peer p to a list of peers 
(create-map* (lambda (tp p) `(,p SND ,(cdr tp) GRAFT ,(car tp)))
             lotopicpeerp
             loevp
             (:name mk-grafts)
             (:fixed-vars ((peerp p))))

(check= (map* mk-grafts '((FM . A) (DS . B)) 'P)
        '((P SND A GRAFT FM) (P SND B GRAFT DS)))
\end{lstlisting}

Given an admitted function name or a lambda expression \codify{f} of
type \codify{a $\times$ b $\rightarrow$ b}, the higher order function
\codify{create-reduce*} defines a function \codify{reduce*-*f} which
accepts a list of elements of type \codify{a}, an initial accumulator
value of type \codify{b}, and returns a reduction of the list using
\codify{f}, from left to right.
\begin{lstlisting}
  ;; function to extract all the subscribers (neighboring peers) from a topic-lop-map             
  (create-reduce* (lambda (tp-ps tmp) (app tmp (cdr tp-ps)))
                  lopp
                  topic-lop-mapp
                  (:name subscribers))

  (check= (reduce* subscribers '()
                   '((T1 P1 P2 P3)
                     (T2 P4 P5 P1)))
          '(P4 P5 P1 P1 P2 P3))
\end{lstlisting}
\section{Reasoning about the scoring function}
\label{sec:reasoning}
Based on the observation that honest peers can be distinguished from
malicious ones based on their observable behaviors (using local
counters and scores), and thus, the overall network can be made more
secure and performant if every honest peer promotes their
well-behaving neighbors and demotes poorly-behaved ones, we came up
with the following informal fundamental property.

\textbf{Fundamental Property of \gs\ Defense Mechanisms.} \emph{Peers
  who behave poorly will be demoted by their neighbors.  Peers who
  behave better-than-average will be promoted by their neighbors.
  Promotion/demotion is entirely based on local peer behavior.}

\begin{figure}[H]
  \centering
  \tikzset{
  hex/.pic={
    \newdimen\R
    \R=0.3cm
    \filldraw (0:\R)
    \foreach \x in {60,120,...,360} {  -- (\x:\R) };
    \foreach \x in {60,120,...,360} { \node (-\x) at (\x:\R/2){}; };
    \node (-c) at (0,0) {};
    \node (-l) at (30:1.6*\R) {#1};
    \foreach \x in {60,180,300} { \draw[color=white] (0,0) -- (\x:\R); };
  }}

\begin{tikzpicture}
  
  \begin{scope}[fill opacity=0.2]
    \node[ellipse,
    draw=none,
    fill = RoyalBlue!70,
    minimum height = 2cm,
    minimum width = 5cm] (e) at (1,0.5) {};
    \node[ellipse,
    draw=none,
    fill = Red!40,
    minimum height = 2cm,
    minimum width = 4cm] (e) at (2.8,0.5) {};
    \node[ellipse,
    draw=none,
    fill = Green!40,
    minimum height = 2cm,
    minimum width = 2cm] (e) at (-1.5,0.5) {};
  \end{scope}

  \pic[Cerulean, rotate=90] (c) {hex=C};
  \pic[Cerulean, rotate=90] (d) at (0,1) {hex=D};
  \pic[Purple, rotate=90] (b) at (2,0.5) {hex=B};
  \pic[Purple, rotate=90] (a) at (3,0.5) {hex=A};
  \pic[Red, rotate=90] (e) at (4,0.5) {hex=E};
  \pic[BlueGreen, rotate=90] (f) at (-1,0.25) {hex=F};
  \pic[ForestGreen, rotate=90] (g) at  (-2,0.75) {hex=G};
  
  \begin{scope} [on background layer]
  \draw [color=Cerulean,line width=0.5mm] (c-c) -- (b-c);
  \draw [color=Cerulean,line width=0.5mm] (b-c) -- (d-c);
  \draw [color=Cerulean,line width=0.5mm] (c-c) -- (f-c);
  \draw [color=red,line width=0.5mm] (a-c) -- (e-c);
  \draw [color=Cerulean,line width=0.5mm] (a-c) -- (b-c);
  \draw [color=Red,line width=0.5mm] (a-60) -- (b-60);
  \draw [color=ForestGreen,line width=0.5mm] (f-c) -- (g-c);
  \draw [color=ForestGreen,-latex,line width=0.5mm] (d-c) -- (f-300);
  \draw [color=ForestGreen,-latex,line width=0.5mm] (d-c) -- (g-360);
\end{scope}
\end{tikzpicture}
  \caption{An example network.}
  \label{fig:example}
\end{figure}
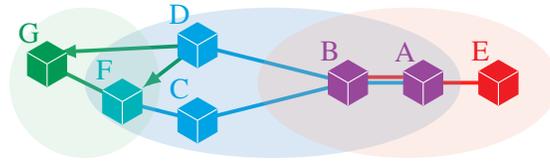
Before reasoning about the formalization of this fundamental property
later in the paper, we discuss the importance of this fundamental
property. Consider the simple example shown in
Figure~\ref{fig:example}, where peers A and B subscribe to, and are
mesh neighbors in both Red and Blue topics. It might be possible for B
to observe that A behaves perfectly well in the Blue topic while
simultaneously misbehaving in the Red topic. This is not good for B
because it depends on A for all of its messages in the Red topic. Note
that this is a very simplified example. In an actual network, B could
have several other neighbors subscribed to the Red topic. But as we will
show later, it is equally trivial to have a scenario where all of B's
neighbors isolate it from communications in the Red topic. In this
example, we want B to prune A from its Red mesh in hopes of finding a
better mesh neighbor later on. Reasoning about this fundamental
property directly would be difficult due to the massive search-space
of possible attack vectors. Hence, we focus on the following liveness
property capturing the essence of the fundamental property:
\begin{property}\label{p1}
  \emph{If a peer's score relating to its performance in any topic is
    continuously non-positive, then the peer's overall score should
    eventually be non-positive:}
  \[\forall q,\tau::\langle
  \glb(\mathit{score(q)}\textit{ for topic }\tau \leq 0) \Rightarrow
  \fin(\mathit{score(q)} \leq 0) \rangle \]
\end{property}  
\emph{where score$(q)$ for topic $t$ is defined below.}
\[
  \mbox{tw}^\tau \times \sum_{i \in \{1,2,3,3b,4\}} w_i^\tau P_i^\tau
\] 

Notice that Property~\ref{p1} is temporal. We write the non-temporal
version of this property in context of \Eth\ (using \Eth\
\codify{twp}) in ACL2s as shown below, and disprove the temporal version
using an induction argument later in the paper.
\begin{property} \normalfont
    \label{p1acl2s} \hfill
\begin{lstlisting}
(property (ptc :pt-tctrs-map pcm :p-gctrs-map p :peer top :topic)
  :hyps  (^ (member-equal `(,p . ,top) (acl2::alist-keys ptc))
            (> (lookup-score p (calc-nbr-scores-map ptc pcm *eth-twp*)) 0))
  (> (calcScoreTopic (lookup-tctrs p top ptc) (mget top *eth-twp*)) 0))
\end{lstlisting}
\end{property}

The following is one of the counter-examples to the above property,
generated by cgen in \acls:
            
\begin{lstlisting}[label=ctrex]
((top 'agg)
 (p 'p4)
 (pcm '((p3449 (:0tag . gctrs) (:apco . 0) (:bhvo . 0) (:ipco . 0))
        (p3450 (:0tag . gctrs) (:apco . 0) (:bhvo . 0) (:ipco . 0))
        (p3451 (:0tag . gctrs) (:apco . 0) (:bhvo . 0) (:ipco . 0))))
 (ptc '(((p4 . agg)
         (:0tag . tctrs)
         (:firstmessagedeliveries . 0)
         (:invalidmessagedeliveries . 0)
         (:meshfailurepenalty . 0)
         (:meshmessagedeliveries . 1)
         (:meshtime . 42))
        ((p4 . blocks)
         (:0tag . tctrs)
         (:firstmessagedeliveries . 324)
         (:invalidmessagedeliveries . 0)
         (:meshfailurepenalty . 0)
         (:meshmessagedeliveries . 330)
         (:meshtime . 377))
        ((p4 . sub1)
         (:0tag . tctrs)
         (:firstmessagedeliveries . 371)
         (:invalidmessagedeliveries . 0)
         (:meshfailurepenalty . 0)
         (:meshmessagedeliveries . 377)
         (:meshtime . 324))
        ((p4 . sub2)
         (:0tag . tctrs)
         (:firstmessagedeliveries . 318)
         (:invalidmessagedeliveries . 0)
         (:meshfailurepenalty . 0)
         (:meshmessagedeliveries . 324)
         (:meshtime . 371))
         ... ))
\end{lstlisting}
For brevity, we omit entries for peer-topic key values for peers other
than \codify{p4}. In property-based testing, the free variables of a property under test
are assigned values using a synergistic combination of theorem proving
and random assignments computed using type-based enumerators
(generators) in an effort to discover counterexamples to the property.
Observe that Property~\ref{p1acl2s} depends on \codify{ptc} and
\codify{pcm} which do not have trivial types. These are maps
containing records which themselves consist of several numerical
values, which makes the search space of possible counter-examples
immensely large. In order to make it easier for cgen to find
counter-examples, we wrote custom enumerators to enumerate restricted
values for \codify{topic}, \codify{tctrs}, \codify{gctrs},
\codify{pt-tctrs-map} and \codify{p-gctrs-map}. Specifically, we limit
the penalties on the score due to some counters, such that the
negative contributions due to misbehavior are comparable to the
positive contributions due to good behavior. Below we define custom
enumerators for \codify{topic} and \codify{tctrs}.

\begin{lstlisting}
(definec topics () :tl
  ;; valid topics used in Ethereum
  '(AGG BLOCKS SUB1 SUB2 SUB3))

(definec nth-topic-custom (n :nat) :symbol
  (nth (mod n (len (topics))) (topics)))
(defdata lows (range integer (0 <= _ <= 1))) ;; high values
(defdata-subtype lows nat)
(defdata highs (range integer (300 < _ <= 400))) ;; low values
(defdata-subtype highs nat)

(defun nth-bad-counters-custom (n)
;; setting invalidMessageDeliveries and meshFailurePenalty to 0 due to high penalty
(defun nth-good-counters-custom (n)
  (tctrs 0 (nth-highs (+ n 2)) (nth-highs (+ n 3)) (nth-highs (+ n 4)) 0))

(defun nth-counters-custom (n) ;; custom enumerator for tctrs
  (if (== 0 (mod n 4))
      (nth-bad-counters-custom n)
    (nth-good-counters-custom n)))
  \end{lstlisting}

  Besides Property~\ref{p1acl2s}, we formalize three safety properties
  for GossipSub, stated below. Together, these four are the most
  general properties of the score function, which must hold in order
  for the fundamental property to hold.
\begin{property}
\label{p2}
\emph{Increasing bad-performance counters (which are multiplied with
  negative weights) should decrease the overall score.}
\end{property}
\begin{property}
\label{p3}
\emph{Increasing good-performance counters (which are multiplied with
  positive weights) will not decrease the score for a mesh peer that has
  been in the mesh for a sufficiently long time.}
\end{property}
\begin{property}
\label{p4}
\emph{If two peers subscribe to the same topics, and achieve identical
  per-topic counters, and identical global counters, then they achieve
  identical scores.}
\end{property}
We are able to find counterexamples to Property~\ref{p2} in much the
same way as~\ref{p1acl2s}, however, in this work we focus on the
counterexamples to Property~\ref{p1acl2s}, which are more interesting
in terms of attack generation.  We manually prove that
Property~\ref{p3} holds over all configurations (available with the
paper artifacts).  The proof that Property~\ref{p4} holds over all
configurations follows directly from referential transparency.

We also prove a limit on the maximum score achievable in a topic, as
shown below.
  \begin{lstlisting}[label=prop:max]
(property max-topic-score (tctrs :tctrs weights :weights params :params)
  (<= (calcScoreTopic tctrs (cons weights params))
      (* (params-topicweight params)
         (+ (* (mget :w1 weights) (params-timeQuantaInMeshCap params))
            (* (mget :w2 weights) (params-p2cap params))))))
\end{lstlisting}

\section{Attack Generation}
\label{sec:attack}
The counter-example~\ref{ctrex} which we obtained in the previous
section is a specification for an unsafe state that does not satisfy
Property~\ref{p1acl2s} for an \Eth\ network. However, we are also
interested in characterizing and generating attacks against an \Eth\
\gs\ network for three main reasons: (1) to show that an unsafe state
is \textbf{reachable} from a reasonable start state, (2)
\textbf{invalidation } of our temporal Property~\ref{p1} using the
trace generated from the attack, and finally (3) demonstration of
\textbf{scalability} of our attack to large networks, typically of the
size and shape used by real world applications.

\begin{figure}
  \centering
  \tikzset{
  hex/.pic={
    \newdimen\R
    \R=0.3cm
    \filldraw (0:\R)
    \foreach \x in {60,120,...,360} {  -- (\x:\R) };
    \foreach \x in {60,120,...,360} { \node (-\x) at (\x:\R/2){}; };
    \node (-c) at (0,0) {};
    \node (-l) at (30:1.6*\R) {#1};
    \foreach \x in {60,180,300} { \draw[color=white] (0,0) -- (\x:\R); };
  }}

\begin{tikzpicture}
  
  \begin{scope}[fill opacity=0.2]
    \node[ellipse,
    draw=none,
    fill = RoyalBlue!70,
    minimum height = 2cm,
    minimum width = 3cm] (e) at (2,0.5) {};
    \node[ellipse,
    draw=none,
    fill = Red!40,
    minimum height = 2cm,
    minimum width = 3.5cm] (e) at (2.8,0.5) {};
    \node[ellipse,
    draw=none,
    fill = Yellow!80,
    minimum height = 2cm,
    minimum width = 2.5cm] (e) at (2.8,0.5) {};
    \node[ellipse,
    draw=none,
    fill = Green!40,
    minimum height = 2cm,
    minimum width = 3.5cm] (e) at (2.8,0.75) {};
  \end{scope}

  \pic[Purple, rotate=90] (b) at (1.8,0.5) {hex=$A$};
  \pic[Purple, rotate=90] (a) at (3,0.5) {hex=$V$};
  
  \begin{scope} [on background layer]
  \draw [dotted,color=Cerulean,line width=0.5mm] (a-c) -- (b-c);
  \draw [dotted,color=Red,line width=0.5mm] (a-60) -- (b-60);
  \draw [color=Yellow,line width=0.5mm] (a-240) -- (b-240);
  \draw [color=Green,line width=0.5mm] (a-180) -- (b-180);
\end{scope}
\end{tikzpicture}
  \caption{An example \ag{2} attack gadget $\langle A, V, \{\mbox{Red}, \mbox{Blue}\} \rangle$.}
  \label{fig:gadget}
\end{figure}
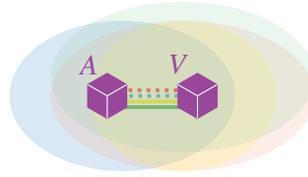

Counter-example~\ref{ctrex} suggests that an unsafe state is one where
a neighboring peer throttles communication in a particular topic
while maintaining an overall positive score, hence, avoiding getting
pruned. Using this insight, we design attack gadgets that can
perpetrate such attacks locally. We define an \emph{attack gadget} as a tuple
$\langle A, V, S \rangle$, where $A, V$ are peers ($A$ is the attacker
and $V$ is the victim), $S$ is a set of subnet topics (the attacked
topics), and $A, V$ are mesh neighbors over a set of topics that is a
superset of $S$. For each $i \in \mathds{N}$, we define \ag{i} to be
the set of attack gadgets where $|S|=i$. Figure~\ref{fig:gadget}
illustrates an example attack gadget in \ag{2} where peers $A$ and $V$
are neighbors in four meshes corresponding to topics: Red, Yellow,
Blue and Green, out of which $A$ is attacking $V$ in
$S=\{\mbox{Red}, \mbox{Blue}\}$.
We generate a sequence of events consisting of message transmission
events from $A$ to $V$ (referred to as \codify{a} and \codify{v} in
code) as well as heart-beat events at $V$ ($V$ updates the scores of
its peers during heart--beat events). These events are designed to
either restrict or completely block communication to $V$ in the
attacked topics while maintaining normal communication rate in all the
other topics, as shown in the following code snippet.
  \begin{lstlisting}
(definecd emit-evnts (a v :peer ts ats :lot n m e :nat) :loev
  ;; mesh message deliveries in attacked topics 
  (app (emit-meshmsgdeliveries-peer-topics a v ats m)
       ;; mesh message deliveries in other topics 
       (emit-meshmsgdeliveries-peer-topics a v (set-difference-equal ts ats) n)
       ;; heart-beat events at the victim node
       `((,p2 HBM ,e))))
\end{lstlisting}
The expression \codify{(emit-meshmsgdeliveries-peer-topics a v ts ats
  n m e)} generates a list of events $E$ sending \codify{m} mesh
messages in the attacked topics and \codify{n} mesh messages in the
other topics from peer \codify{a} to peer \codify{v} per heart-beat at
\codify{v} which happens every \codify{e} seconds. \codify{m} is
generally set to \codify{0} or \codify{1} in order to block or
throttle communication in the attacked topic meshes. We choose a start
state of the network based on actual full-network topologies of the
\Eth\ testnets Ropsten (shown in Figure~\ref{fig:network}), Goerli and
Rinkeby, as measured by Li
et. al~\cite{toposhot}. Table~\ref{tab:eth-topos} characterizes each
of these topologies.\\

\begin{table}[h]
\begin{center}
\begin{tabular}{|l|r|r|r|r|r|}
  \hline
  \multirow{2}{*}{Network} &
  \multirow{2}{*}{Nodes} &
  \multicolumn{3}{c|}{Degree} &
  \multirow{2}{*}{Diameter} \\
  \cline{3-5}
  && min & max & avg & \\
  \hline
  Ropsten  & 588   & 1  & 418   & 25.49  &  5 \\ \hline
  Goerli   & 1355  & 1  & 712   & 28.26  & 5  \\ \hline
  Rinkeby  & 446   & 1  & 191   & 68.96  & 6  \\ \hline
\end{tabular}
\end{center}
  \caption{\Eth\ Network Characteristics}
  \label{tab:eth-topos}
\end{table}

We create our start state as a \codify{Group}, using the topologies
provided. In this \codify{Group}, we initialize our attack gadget and
simulate its run over the generated sequence of events $E$ using the
\codify{run-network} function. We use function
\codify{scorePropViolation} described below, to
detect violations of Property~\ref{p1acl2s} in peer states occurring
in the output trace.

\begin{lstlisting}
  ;; ats is the list of topics being attacked
  (definec scorePropViolation (ps :peer-state p :peer ats :lot twpm :twp) :boolean
    (match ats
      (() (> (lookup-score p (calc-nbr-scores-map (peer-state-nbr-tctrs ps)
                                                  (peer-state-nbr-gctrs ps) twpm))
             0))
      ((top . rst) (^ (< (calcScoreTopic
                          (lookup-tctrs p top (peer-state-nbr-tctrs ps))
                          (mget top twpm))
                         0)
                      (scorePropViolation ps p rst twpm)))))
\end{lstlisting}

We observe that after the first heart-beat event at the victim peer,
the state of the network at each subsequent heart-beat is
identical. Since Property~\ref{p1acl2s} is being violated at each of
these events, we make an inductive claim that it will forever be
violated, thus giving a counter-example to our liveness
Property~\ref{p1}. We wrote optimized versions of the
\codify{run-network} function to generate traces of property
violations as a list of booleans, instead of generating the full trace
of type \codify{egl}. We ran our simulations for 100,000 events to
ensure that we ended up in identical states, taking about a
minute on an M1 Macbook Air.

Finally, we scaled our attacks to build \emph{eclipse} and
\emph{network partition} attacks using a combination of attack
gadgets. \Cref{fig:eclipse,fig:partition} give
an intuition of how to combine our attack gadgets to carry out these
attacks. Our companion paper discusses the specifics of these attacks
in more detail.
\begin{figure}[!t]
    \centering
    \begin{minipage}{.5\textwidth}
        \centering
        \tikzset{
  hex/.pic={
    \newdimen\R
    \R=0.3cm
    \filldraw (0:\R)
    \foreach \x in {60,120,...,360} {  -- (\x:\R) };
    \foreach \x in {60,120,...,360} { \node (-\x) at (\x:\R/2){}; };
    \node (-c) at (0,0) {};
    \node (-l) at (30:1.6*\R) {#1};
    \foreach \x in {60,180,300} { \draw[color=white] (0,0) -- (\x:\R); };
  }}

\begin{tikzpicture}
  
  \begin{scope}[fill opacity=0.2]
    \node[ellipse,
    draw=none,
    fill = RoyalBlue!70,
    minimum height = 5cm,
    minimum width = 5cm] (e) at (2,0.5) {};
    \node[ellipse,
    draw=none,
    fill = Red!40,
    minimum height = 5cm,
    minimum width = 5.5cm] (e) at (2.8,0.5) {};
    \node[ellipse,
    draw=none,
    fill = Yellow!80,
    minimum height = 5cm,
    minimum width = 5.5cm] (e) at (2.8,0.5) {};
    \node[ellipse,
    draw=none,
    fill = Green!40,
    minimum height = 5cm,
    minimum width = 5.5cm] (e) at (2.8,0.75) {};
  \end{scope}

  \pic[Purple, rotate=90] (v) at (3,0.5) {hex=$V$};
  \pic[Purple, rotate=90] (a2) at (1.5,1.15) {hex=$A^2$};
  \pic[Purple, rotate=90] (a3) at (3.5,1.25) {hex=$A^3$};
  \pic[Purple, rotate=90] (a4) at (2.2,-0.40) {hex=$A^4$};
  \pic[Purple, rotate=90] (a1) at (0.8,0.5) {hex=$A^1$};
  
  \begin{scope} [on background layer]
  \draw [dotted,color=Cerulean,line width=0.5mm] (v-c) -- (a1-c);
  \draw [dotted,color=Red,line width=0.5mm] (v-60) -- (a1-60);
  \draw [color=Yellow,line width=0.5mm] (v-240) -- (a1-240);
  \draw [color=Green,line width=0.5mm] (v-180) -- (a1-180);
    \draw [dotted,color=Cerulean,line width=0.5mm] (v-c) -- (a2-c);
  \draw [dotted,color=Red,line width=0.5mm] (v-60) -- (a2-60);
  \draw [color=Yellow,line width=0.5mm] (v-240) -- (a2-240);
  \draw [color=Green,line width=0.5mm] (v-180) -- (a2-180);
      \draw [dotted,color=Cerulean,line width=0.5mm] (v-c) -- (a3-c);
  \draw [dotted,color=Red,line width=0.5mm] (v-60) -- (a3-60);
  \draw [color=Yellow,line width=0.5mm] (v-240) -- (a3-240);
  \draw [color=Green,line width=0.5mm] (v-180) -- (a3-180);
      \draw [dotted,color=Cerulean,line width=0.5mm] (v-c) -- (a4-c);
  \draw [dotted,color=Red,line width=0.5mm] (v-60) -- (a4-60);
  \draw [color=Yellow,line width=0.5mm] (v-240) -- (a4-240);
  \draw [color=Green,line width=0.5mm] (v-180) -- (a4-180);
\end{scope}
\end{tikzpicture}
  \caption{An eclipse attack using $AG_2$ gadgets}
  \label{fig:eclipse}
    \end{minipage}%
    \begin{minipage}{0.5\textwidth}
        \centering
         \tikzset{
  hex/.pic={
    \newdimen\R
    \R=0.3cm
    \filldraw (0:\R)
    \foreach \x in {60,120,...,360} {  -- (\x:\R) };
    \foreach \x in {60,120,...,360} { \node (-\x) at (\x:\R/2){}; };
    \node (-c) at (0,0) {};
    \node (-l) at (30:1.6*\R) {#1};
    \foreach \x in {60,180,300} { \draw[color=white] (0,0) -- (\x:\R); };
  }}

\begin{tikzpicture}
  
  \begin{scope}[fill opacity=0.2]
    \node[ellipse,
    draw=none,
    fill = RoyalBlue!70,
    minimum height = 5cm,
    minimum width = 5cm] (e) at (2,1.5) {};
    \node[ellipse,
    draw=none,
    fill = Red!40,
    minimum height = 5cm,
    minimum width = 5.5cm] (e) at (2.8,1.5) {};
    \node[ellipse,
    draw=none,
    fill = Yellow!80,
    minimum height = 5cm,
    minimum width = 5.5cm] (e) at (2.8,1.5) {};
    \node[ellipse,
    draw=none,
    fill = Green!40,
    minimum height = 5cm,
    minimum width = 5.5cm] (e) at (2.8,1.75) {};
  \end{scope}

  \pic[Purple, rotate=90] (v1) at (3,0.5) {hex=$V^1$};
  \pic[Purple, rotate=90] (v2) at (2.8,3.7) {hex=$V^2$};
  \pic[Purple, rotate=90] (v3) at (3.2,2.6) {hex=$V^3$};
  \pic[Purple, rotate=90] (v4) at (3.1,1.5) {hex=$V^4$};
  \pic[Purple, rotate=90] (a2) at (1,1.5) {hex=$A^2$};
  \pic[Purple, rotate=90] (a4) at (1,3.0) {hex=$A^3$};
  \pic[Purple, rotate=90] (a1) at (0.8,0.5) {hex=$A^1$};
  
  \begin{scope} [on background layer]
  \draw [dotted,color=Cerulean,line width=0.5mm] (v1-c) -- (a1-c);
  \draw [dotted,color=Red,line width=0.5mm] (v1-60) -- (a1-60);
  \draw [color=Yellow,line width=0.5mm] (v1-240) -- (a1-240);
  \draw [color=Green,line width=0.5mm] (v1-180) -- (a1-180);
    \draw [dotted,color=Cerulean,line width=0.5mm] (v1-c) -- (a2-c);
  \draw [dotted,color=Red,line width=0.5mm] (v1-60) -- (a2-60);
  \draw [color=Yellow,line width=0.5mm] (v1-240) -- (a2-240);
  \draw [color=Green,line width=0.5mm] (v1-180) -- (a2-180);
      \draw [dotted,color=Cerulean,line width=0.5mm] (v3-c) -- (a4-c);
  \draw [dotted,color=Red,line width=0.5mm] (v3-60) -- (a4-60);
  \draw [color=Yellow,line width=0.5mm] (v3-240) -- (a4-240);
  \draw [color=Green,line width=0.5mm] (v3-180) -- (a4-180);
      \draw [dotted,color=Cerulean,line width=0.5mm] (v2-c) -- (a4-c);
  \draw [dotted,color=Red,line width=0.5mm] (v2-60) -- (a4-60);
  \draw [color=Yellow,line width=0.5mm] (v2-240) -- (a4-240);
  \draw [color=Green,line width=0.5mm] (v2-180) -- (a4-180);
        \draw [dotted,color=Cerulean,line width=0.5mm] (v4-c) -- (a2-c);
  \draw [dotted,color=Red,line width=0.5mm] (v4-60) -- (a2-60);
  \draw [color=Yellow,line width=0.5mm] (v4-240) -- (a2-240);
  \draw [color=Green,line width=0.5mm] (v4-180) -- (a2-180);
\end{scope}
\end{tikzpicture}
  \caption{A partition attack using $AG_2$ gadgets}
  \label{fig:partition}
    \end{minipage}
\end{figure}

\section{Limitations}
\label{sec:limitations} 
We now discuss limitations of our model. The most
crucial aspect of property-based testing is counter-example generation for invalid
properties. As explained previously, our properties depend on
complex types, for which we had to write custom
enumerators. Coming up with enumerators that had a high
probability of satisfying the hypotheses of our properties required considerable 
analysis of \Eth\ so as to
restrict certain \codify{tctrs} values from skewing the
scores too much. Testing properties for new applications will likewise require
writing new custom enumerators. One might need to write a new
event generator as well, possibly generating events of shapes different
from the ones we showed.

\section{Related Work}
\label{sec:related}
The Protocol Labs ResNetLab and software audit firm Least Authority
tested \gs\ against a list of specific pre-programmed attack
scenarios~\cite{resNetGS} designed to degrade overall network
performance using a network emulator called
\textsc{Testground}~\cite{testground}. Due to their use of simplified
configurations with only one topic (and because simple testing is not
enough to find bugs) they found that all the attacks failed against
\gs, and the \score\ made \gs\ more resilient to malicious nodes
attacks than the other tested protocols~\cite{vyzovitis2020gossipsub}.
Separate from simulation testing, Least Authority also audited the
\golang\ implementation and provided recommendations for
improvement~\cite{leastAuth20}. Though our work contributes the first
formalization of \gs, there has been considerable previous work on
utilizing formal methods to
reason about distributed systems. We survey such works below.

\noindent
\textbf{Model Checking based approaches.}  Lamport's modeling language
\textsc{TLA+}~\cite{lamport2002specifying} and the corresponding TLC
model checker~\cite{yu1999model} have been used to analyze properties
of distributed systems including \textsc{Disk
  Paxos}~\cite{gafni2003disk},
\textsc{MongoRaftReconfig}~\cite{schultz2022formal}, Byzantine
\textsc{Paxos}~\cite{lamport2011byzantizing},
\textsc{Spire}~\cite{koutanov2021spire}, etc. McMillan and Zuck
applied specification-based testing to the \textsc{Quic} protocol, and
found vulnerabilities~\cite{quicfmsigcomm19}.  Wu~et.~al. formally
modeled the Bluetooth stack using \textsc{ProVerif}, a model checker,
and found five known vulnerabilities and two new
ones~\cite{bluetoothfmsp2022}.  Chothia~et.~al. demonstrated the use
of \textsc{ProVerif} to verify distance-bounding protocols, \eg, those
used by MasterCard and NXP~\cite{chothia2018modelling}.  Separately,
Chothia modeled the \textsc{MUTE} anonymous file-sharing system using
the $\pi$-calculus, and proved the system insecure (discovering a novel
attack)~\cite{chothia2006analysing}.  Cremers~et.~al. modeled all
handshake modes of TLS~1.3 using \textsc{Tamarin}, another model
checking tool, and discovered an unexpected
behavior~\cite{tls13fmccs2017}. An issue with using model checking
tools like ProVerif or Tamarin to verify a protocol like \gs\ is the
immense size of the state space needed to be checked, making them
infeasible for our use.
 
\noindent
\textbf{Refinement-based proof formalization.} The theory of
refinement has proved to be useful for enabling the mechanical
verification of distributed systems' properties. Manolios' work on
refinement~\cite{pete-dissertation,ManoliosNS99,Manolios03} has been
previously used for mechanical verification of pipelined
processors~\cite{DBLP:conf/fmcad/Manolios00,DBLP:conf/date/ManoliosS04,DBLP:conf/date/ManoliosS05,DBLP:journals/tvlsi/ManoliosS08,DBLP:journals/todaes/ManoliosS08}. Manolios~et.~al. combined
theorem proving (using refinement maps) with model checking to verify
the alternating bit
protocol.~\cite{ManoliosNS99}. \textsc{IronFleet}~\cite{hawblitzel2015ironfleet}
refines \textsc{TLA} style state-machine specification of a
\textsc{Paxos}-based library and a sharded key-value store to low
level implementation in Dafny (a SMT based program verifier) for
verification using Hoare-logic. Woo~et.~al.~\cite{woos2016planning}
formally verified 90 properties of the \textsc{RAFT}~protocol using
\textsc{Verdi}~\cite{wilcox2015verdi}, a tool they built in the
\textsc{Coq} proof assistant. Though they did not build an executable
model, their framework can be used to extract an executable protocol
implementation in OCaml. \textsc{Verdi} provides verified \emph{system
  transformers} used to refine a system in an ideal fault model to a
more realistic fault model, without any proof overhead on part of the
user. We believe that looking at \gs\ through the lens of refinement
will be interesting because, not only will it allow us to explain why
it failed our properties, but also guide us towards improving it.

\noindent
\textbf{Inductive-invariant based proof formalization.}
Padon~et.~al.~\cite{padon2017paxos} proved the correctness of a simple
model of Paxos described in Effectively Propositional Logic (a
decidable fragment of First Order Logic) using
\textsc{IVY}~\cite{padon2016ivy}, a SMT-based safety verification
tool.  \textsc{IVY} can be used for verifying inductive invariants
about global states of a distributed protocol. Both the modeling and
the specification languages of \textsc{IVY} are restricted to a
decidable fragment of First Order Logic to ensure that all
verification conditions can be checked algorithmically. Hippel
et. al.~\cite{KarnFormal} also used IVY to formally describe and
reason about Karn's Algorithm, a mechanism used to study rount trip
times of message transmissions. However, since IVY lacks a theory of
rationals, modelling the scoring function of \gs\ would not have been
possible using this tool.

\noindent
\textbf{Full stack verification.}
Certain high-assurance distributed systems might require the whole
stack to be formally-verified. Such applications could, for instance,
be implemented on top of \textsc{sel4}: a high-performance operating
system microkernel that was formally verified against an abstract
specification using higher-order logic~\cite{klein2009sel4}. Another
example is the fully verified \emph{CLI stack}~\cite{cli}, a system
comprising of an operating system with some applications running in
it, operational semantics for two high level languages, a stack based
assembly language and the instruction set architecture (ISA), all the
way down to the register transfer level (RTL) design for a
microprocessor. The full stack was verified in Nqthm~\cite{nqthm}.

\section{Conclusion and Future Work}
\label{sec:conclusion}
In this paper, we described the \gs~protocol, as well as our official
formalization based on its prose specification using the \acls~theorem
prover. We explained our state models, transition functions as well as
design decisions. We showed our security property for \gs\ and how we
were able to find counter-examples against it. Finally, we described
several kinds of attacks we synthesized based on our attack gadgets,
using the couter-examples as specifications.

In the future, we would like to characterize an ideal variant of \gs\
as a refinement of simpler protocols so as to prove safety properties,
as well as to contrast the ideal variant with our current model in
order to better explain why it is susceptible to attacks from
misbehaving peers. We would also like to support reasoning for the
application layer on top of our network model layer, since interesting
bugs can be found at the interface of these two layers.

\nocite{*}
\bibliographystyle{eptcs}
\bibliography{main.bib}
\end{document}